\documentclass[a4paper,11pt]{article}
\pdfoutput=1 

\usepackage{jinstpub} 
                     
\usepackage{listings}

\lstdefinelanguage{JavaScript}{
  keywords={break, case, catch, continue, debugger, default, delete, do, else, false, finally, for, function, if, in, instanceof, new, null, return, switch, this, throw, true, try, typeof, var, void, while, with},
  morecomment=[l]{//},
  morecomment=[s]{/*}{*/},
  morestring=[b]',
  morestring=[b]",
  showstringspaces=false,
  ndkeywords={class, export, boolean, throw, implements, import, this},
  keywordstyle=\color{blue}\bfseries,
  ndkeywordstyle=\color{darkgray}\bfseries,
  identifierstyle=\color{black},
  commentstyle=\color{purple}\ttfamily,
  stringstyle=\color{red}\ttfamily,
  sensitive=true
}

\lstdefinelanguage{Template}{
  keywords={include, import},
  morestring=[b]\|,
  morestring=[b]\|\|,
  showstringspaces=false,
  morecomment=[s]{'}{'},
  morecomment=[s]{"}{"},
  keywordstyle=\color{blue}\bfseries,
  ndkeywordstyle=\color{darkgray},
  identifierstyle=\color{black},
  commentstyle=\color{red}\ttfamily,
  stringstyle=\color{blue}\ttfamily\bfseries,
  sensitive=true
}

\title{Management system for the SND experiments}

\author[1]{K. Pugachev\note{Corresponding author.},}
\author{A. Korol}

\affiliation{Budker Institute of Nuclear Physics, SB RAS,\\ Novosibirsk, 630090, Russia}
\affiliation{Novosibirsk State University,\\ Novosibirsk, 630090, Russia}

\emailAdd{K.V.Pugachev@inp.nsk.su}

\abstract{A new management system for the SND detector experiments (at VEPP-2000 collider in Novosibirsk) is developed. We describe here the interaction between a user and the SND databases. These databases contain experiment configuration, conditions and metadata.
The new system is designed in client-server architecture. It has several logical layers corresponding to the users roles. A new template engine is created.
A web application is implemented using Node.js framework. At the time the application provides: showing and editing configuration; showing experiment metadata and experiment conditions data index; showing SND log (prototype).}


\arxivnumber{1705.04317}

\collaboration{\includegraphics[height=17mm]{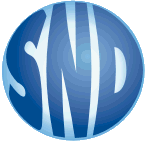}\\[6pt]
  SND collaboration}

\proceeding{The International Conference ``Instrumentation for Colliding Beam Physics'' (INSTR17)\\
  27 February - 3 March, 2017\\
  Novosibirsk, Russia}

\begin{document}
\maketitle
\flushbottom

\section{Introduction}

\begin{figure}[htbp]
\centering
\begin{minipage}{14pc}
\includegraphics[width=14pc]{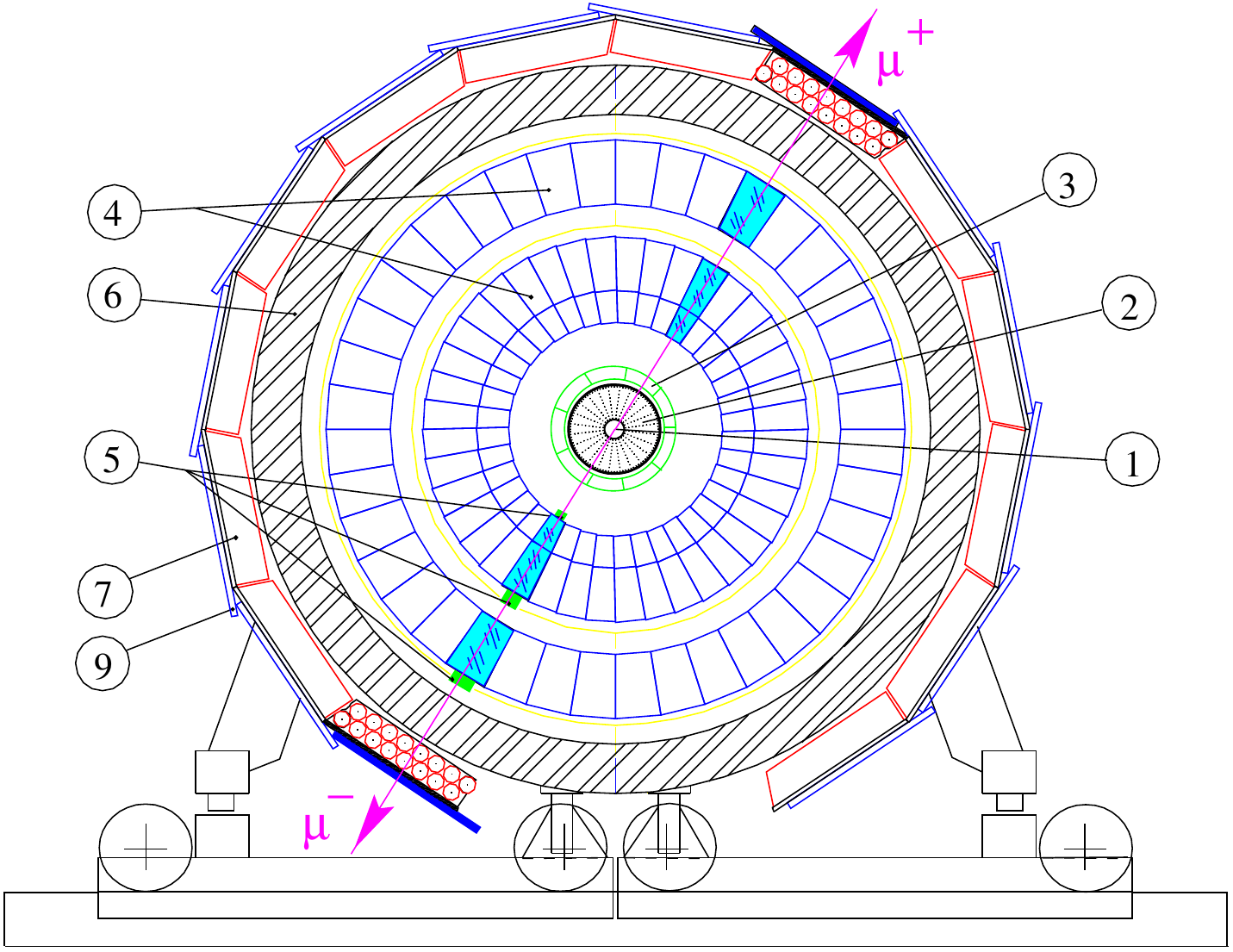}
\end{minipage}\qquad
\begin{minipage}{14pc}
\includegraphics[width=14pc,trim = 0mm 0mm 0mm 14pc, clip]{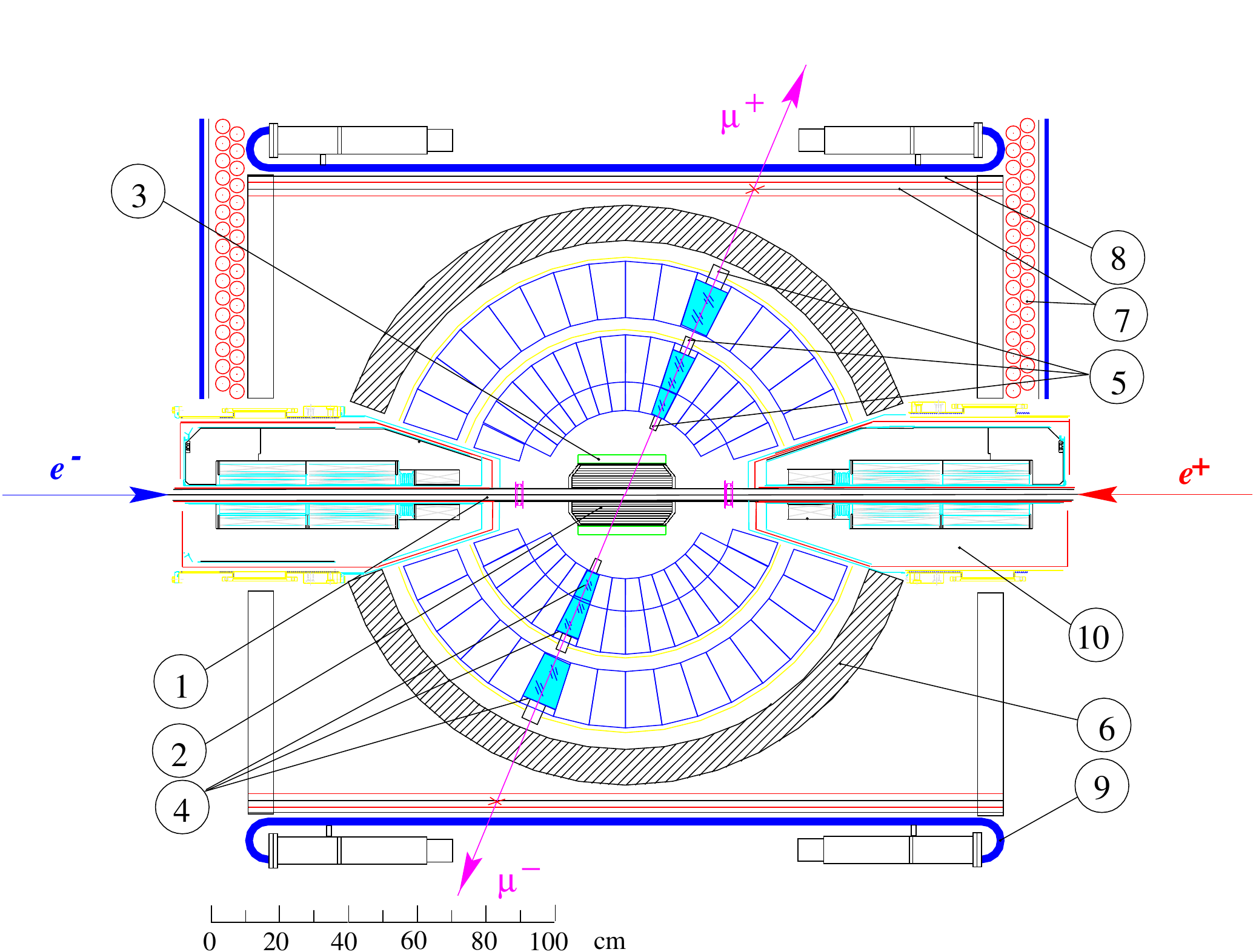}
\end{minipage} 
\caption{\label{SNDpic}The SND detector.}
\end{figure}

The SND detector \cite{SNDAchasov,SNDAbramov,SNDAulchenko} (see figure~\ref{SNDpic}) operates at the VEPP-2000 collider \cite{VEPPKhazin} since 2008. At the time the detector produces hundreds gigabytes of stored information and dozens megabytes of metadata per day. These metadata are used in reconstruction, processing and system control. SND experiments are controlled by operators. SND subsystems are controlled by experts.

The detector has been built in the past century. Its software includes programs, scripts, frameworks, web-servers etc. that cover a lot of experiment tasks. However the current information system is quite old. So it was decided to create a new system that would have some existing features reconsidered and new ones that have not been implemented yet.

\section{The information system implemented}

In order to implement a new information system Node.js framework \cite{NODEJS} has been used. A web-application is created. It allows a user to view and edit experiment configuration; to view experiment conditions data index; to view integral parameters of experiment runs. Some features (showing current SND/VEPP parameters and showing the SND event log) are at the prototype stage.

The SND collaboration consists of 20-30 physicists. It is important to design the system in the way it would require as little programming as possible.
The system user may have some of the following roles: an operator (who monitors the experiments), an expert (who is responsible for some SND subsystems) and a programmer (who develops and supports the information system). An operator may need the experiment information to be properly displayed. An expert needs the same and also can define what information should be displayed. And a programmer can implement new ideas and features at expert or operator request.

We distinguish four roles and created corresponding abstraction layers of the information system:
\begin{itemize}
\item operator (web-interface layer),
\item expert, also template editor (HTML pages templates layer),
\item programmer (template pages variables layer),
\item server developer (server core layer).
\end{itemize}

\section{Template engine}
\subsection{Motivation}

Using a template engine instead of conventional programming language allows experts (who are not required to be programmers) to create or modify some interfaces. Information system deals with database queries that are commonly performed asynchronously in JavaScript. Displayed items are table rows that depend on other table rows etc. All the rows result from chains of database queries. So a template engine should deal with asynchronous computations. It is very useful to have that asynchronously computed table rows values represented as simple as possible.

The existing template engines provide various styles and features. Some engines do not introduce new languages (like Plates \cite{Plates} engine that operates with HTML transformations). Some engines support embedding JavaScript in HTML code (e.g. EJS \cite{EJS}, AJS \cite{AJS}). These ones are great but non-professional programmers may use them improperly making code unclear. Some engines propose their own syntax (e.g. Pug \cite{JADE}, dust.js \cite{dustjs}, Kernel \cite{Kernel}). Several ones don't support asynchronous calls (e.g. Pug, EJS; though they have asynchronous counterparts). Several ones don't introduce asynchronously computed values but may support asynchronous functions or filters.

\subsection{Our engine}

The new template engine is created. It allows a user to define scopes containing values and functions that may be used in templates or may be inherited by other scopes. Values may be constant or lazy. Both values and functions may be calculated either synchronously or asynchronously. Lazy values and functions may be marked as pure ones to use caching when possible.

A scope could be defined in a CommonJS module stored in a file having special file extension (for automatic loading by the server). The module should export a function that takes a set of helpers as an argument. That helpers provide a lot of features e.g. defining a scope (\verb'_.scope'), a lazy value (\verb'_.lazy'), SQL query (\verb'_.sql') etc. It is worth mentioning that the order of defining scopes, functions and values does not matter.

The example below describes scope \verb'greeting' that inherits some definitions from \verb'global' and also specifies function \verb'greet' and values \verb'random', \verb'x' and \verb'name'.

\begin{lstlisting}[language=JavaScript]
    module.exports = function(_) {
      _.scope('greeting', ['global'], {
        random: _.lazy(Math.random, 'sync/impure'),
        x: _.lazy(function(cb) { cb(0.5); }, 'async/pure'),
        greet: function(name){ return 'Hello, ' + name; },
        name: _.SESSION('username')
      });
    };
\end{lstlisting}

Once scope \verb'greeting' is defined, its definitions could be used by a template:

\begin{lstlisting}[language=Template]
    import greeting
    -------
    ||header||
    |#if (equals random x)|
      <b>|greet name|</b>
    |end|
    ||footer||
\end{lstlisting}

Template files also have the special file extension that is used for automatic loading too. Values and functions are enclosed in vertical bars. Those ones that contain HTML are enclosed in double vertical bars to prevent from escaping. There are blocks (like \verb'|#if|') that could be specified along with functions and values. Blocks are more complicated to define so a number of useful ones is predefined.

Templates hide all the differences between synchronous and asynchronous computations. It is permitted to apply a synchronous function to an asynchronously computed value and vice versa without mentioning any callbacks.

The server loads templates and scopes files automatically and detects accidental cyclic dependencies. Since declarative style is supported, expression evaluation order is undefined (except for functions that are evaluated after their arguments and compound values that are evaluated after their components). Function arguments are evaluated strictly but it is possible to define functions in a special way and evaluate arguments manually.

\section{Showing the state}

One important goal of the information system is to show the current parameters in some convenient way. Attention should be paid to the set of parameters to be monitored, the parameters values appearance properties.

The number of parameters observed by an operator should be limited. An operator needs to know some collider parameters (the energies, the currents, etc.) and the detector parameters (luminocity, computers temperature, etc.) including experimental runs parameters (number of events recorded, live time, etc.). We decided to divide parameters in three groups: ``VEPP'' (collider parameters), ``SND'' (detector parameters) and ``RUN'' (experiment run parameters for the SND).

An operator also needs to know the last event happened. So showing the SND log is worth implementing. The log contains entries concerning carrying out experiment and processing events.

\begin{figure}[htbp]
\centering
\begin{minipage}{16pc}
\includegraphics[width=16pc]{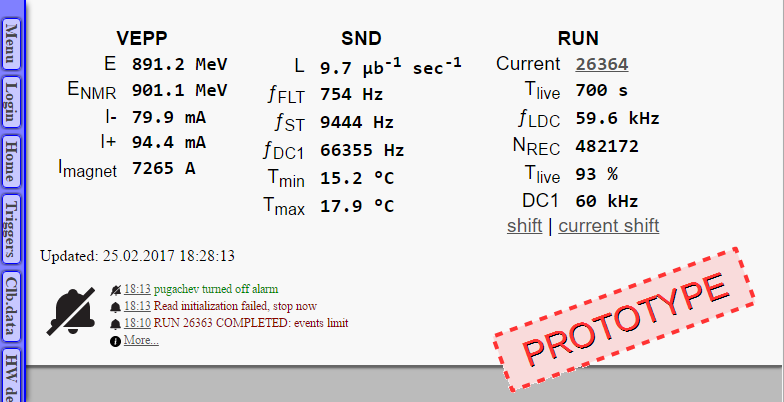}
\end{minipage}\qquad
\begin{minipage}{16pc}
\includegraphics[width=16pc]{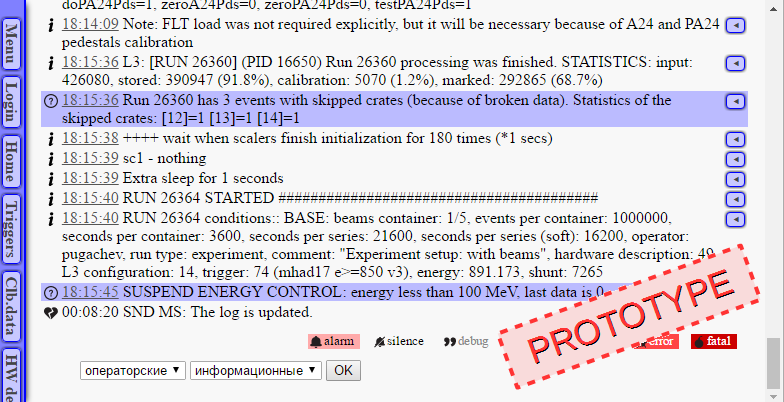}
\end{minipage} 
\caption{\label{AppearanceStateLog}Interface prototypes. Left to right: showing current system state; showing last events.}
\end{figure}

Figure~\ref{AppearanceStateLog} shows prototypes of the mentioned views. The particular details of the interface, such as the parameters precision, relevance period, and highlighting, will be improved later.

Integral parameters of the experiment runs are used by the operator and experts during the experiment.

\begin{figure}[htbp]
\centering
\includegraphics[width=34pc]{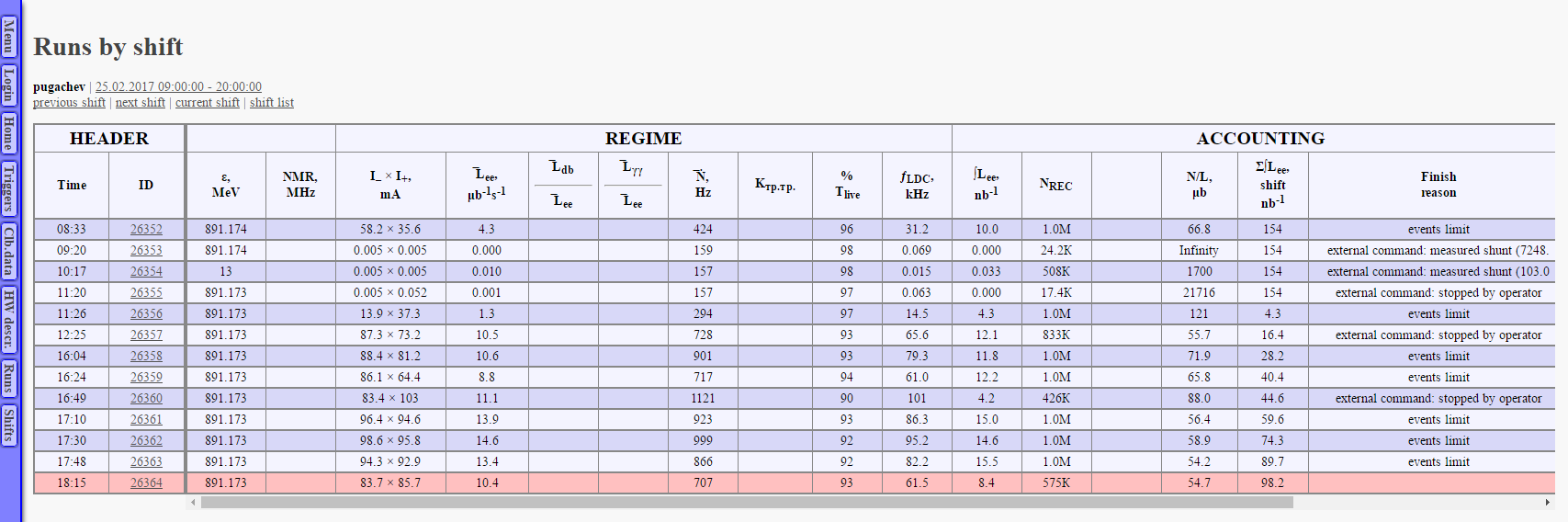}
\caption{\label{AppearanceRuns}Web-interface: showing integral experiment runs parameters for an operator's shift.}
\end{figure}

There is a paper log that embodies a table of the most important runs parameters. It was taken as a basis for a web view (see figure~\ref{AppearanceRuns}).

\section{Conclusion}

A new SND information system is developed. It uses client-server architecture having the four abstraction layers structure designed. The system provides web-interface. It allows a user to deal with some experiment databases and show the most important parameters. HTML pages are rendered using a new template engine.


\acknowledgments

The authors would like to thank other SND team members for help, guiding, discovering bugs and inspiration.
This work is partly supported by the RFBR grants 16-02-00327 and 16-32-00542.


\bibliographystyle{JHEP}
\bibliography{INSTR17-pugachev}{}
\end{document}